\documentclass[aps,prl,reprint,superscriptaddress,nofootinbib]{revtex4-2}
\usepackage{amsmath,amssymb,bm,mathtools}
\usepackage{graphicx}
\usepackage{microtype}
\usepackage[hidelinks]{hyperref}

\newcommand{\Smat}{\mathsf{S}}
\newcommand{\Tmat}{\mathsf{T}}
\newcommand{\Mmat}{\mathsf{M}}
\newcommand{\I}{\mathsf{I}}
\newcommand{\phaseavg}[1]{\left\langle #1\right\rangle_{\phi}}
\newcommand{\dd}{\mathrm{d}}
\newcommand{\order}{\mathcal{O}}

\begin{document}

\title{Geometric Renormalization and a Chirality Threshold in Recursively Coiled Filaments}

\author{Hiroyuki Shima}
\email{hshima@yamanashi.ac.jp}
\affiliation{Department of Environmental Sciences, University of Yamanashi, Kofu, Yamanashi 400-8510, Japan}

\begin{abstract}
Repeated coiling creates a filament hierarchy. We formulate helicalization as an iterated map acting on an arbitrary rod compliance, rather than homogenizing one prescribed construction. A marginal Jordan mode yields an outer-radius inverse-square stiffness class, while pitch disorder creates a Lyapunov threshold between amplified and screened extension--twist response. An exact finite-level rate distinguishes representative amplified and screened cases by level three; direct three-dimensional beam calculations validate the response through level four and convergence through level five.
\end{abstract}

\maketitle

Helices convert one deformation mode into another. This conversion gives coil springs their compliance, couples extension to rotation in filaments, and enables macroscopic chirality to be designed from an ordinary elastic solid \cite{Love1944,Cardou1997,Costello1997,Gomez2025,Frenzel2017}. Recursive helices are themselves established objects: Fletcher \textit{et al.} defined self-similar ``hyperhelices'' and analyzed their wave propagation, while Healey showed that spatial averaging of helical microstructure generates hemitropic rod laws with extension--twist coupling \cite{Fletcher2001,Healey2002}. Static rod and homogenization models have treated prescribed wire ropes, nanotube ropes, and multilevel chiral assemblies, including finite-level stiffness and contact \cite{Elata2004,Usabiaga2008,Zhao2014,Han2023}. Recently, Han \textit{et al.} developed a global--local model for prescribed finite hierarchical chiral helices, including tension--torsion coupling or decoupling and local contact effects \cite{Han2026}. Such hierarchies also underpin artificial muscles and stretchable conductors \cite{Haines2014,Son2019}. Our question is different: we treat helicalization itself as an iterated operator on an arbitrary six-dimensional precursor compliance and ask what spectrum, scaling laws, and random-product behavior emerge with hierarchy depth. A selected finite construction and an iterated constitutive map are therefore non-equivalent problems.

A scalar spring constant cannot resolve these questions. One helicalization generates extension--twist and shear--bending couplings, and these become constitutive inputs to the next level. We instead regard recursive coiling as a renormalization operation on the complete rod compliance. The resulting linear operator has a material-independent spectrum. Its marginal sector generates a universal accumulated-radius law, while its chiral density sector contains a scalar random multiplicative mode with a sharply defined amplification--screening threshold.

\begin{figure*}[t]
\includegraphics[width=0.95\textwidth]{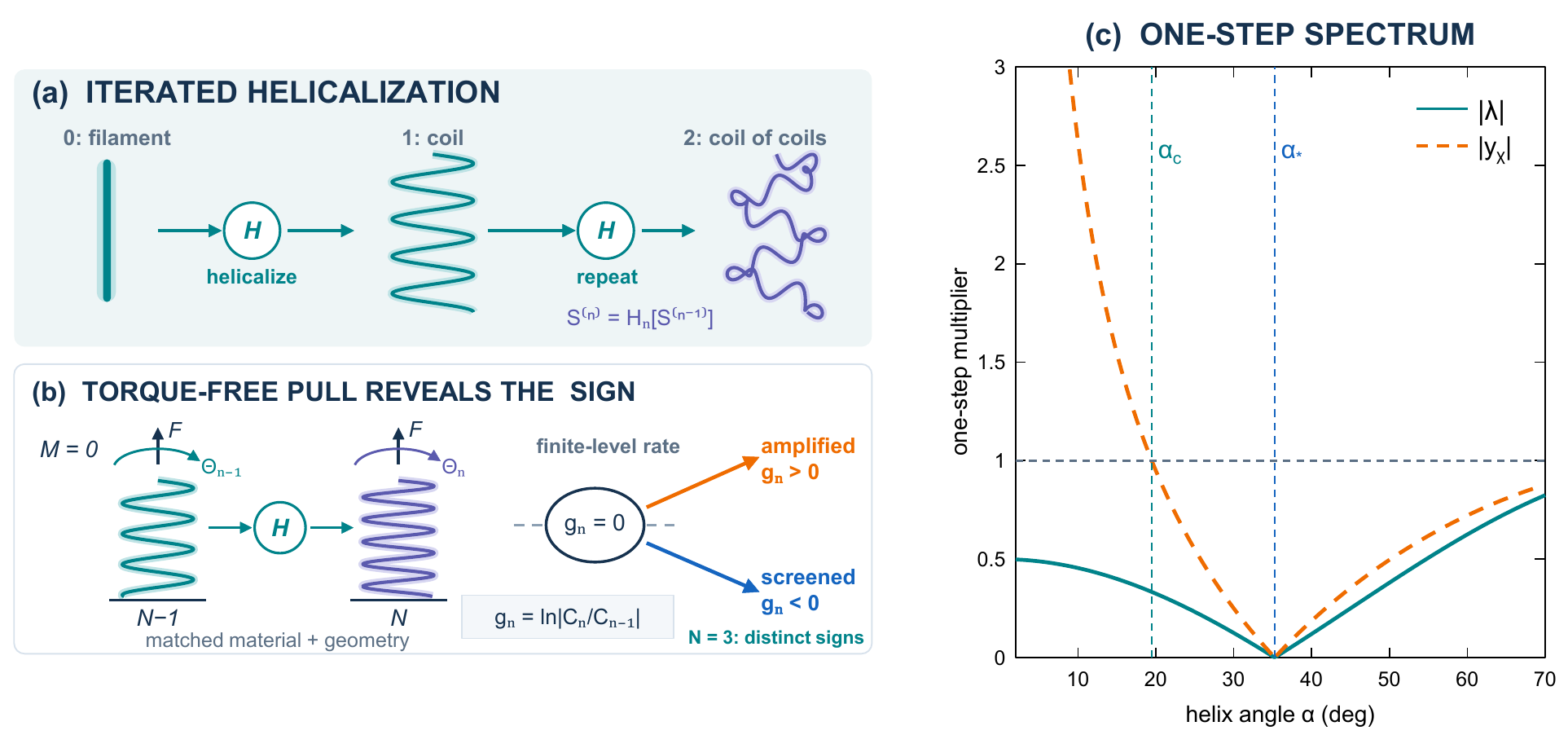}
\caption{\textbf{Geometry, measurement, and renormalization.} (a) Repeated application of the helicalization operator $\mathsf H_n$ creates a nested hierarchy. (b) Matched level-$(N-1)$ and level-$N$ specimens are pulled by axial force $F$ at zero axial torque $M$; their end rotations $\Theta_{N-1}$ and $\Theta_N$ give $\mathcal C_j=-\Theta_j/(FL_jR_j)$ and the finite-step rate $g_N=\ln|\mathcal C_N/\mathcal C_{N-1}|$. Representative amplified and screened cases are distinguishable at $N=3$. (c) One-step multipliers versus helix angle: $|\lambda|$ retains bending--torsion compliance contrast, whereas $|y_\chi|=|\lambda|/\sin\alpha$ multiplies the homogeneous extension--twist coupling-density mode. The line $|y_\chi|=1$ gives $\alpha_c=19.47^\circ$; $\alpha_*=35.26^\circ$ sets $\lambda=0$.}
\label{fig:geometry}
\end{figure*}

The uncoiled filament is level $0$, and an object truncated after $N$ coiling
operations is an $N$-level hierarchy. At level $n=1,\ldots,N$, the axis of the
level-$(n-1)$ object is wound into a circular helix of centerline radius $R_n$,
angle $\alpha_n$ measured from the transverse plane, and handedness
$\chi_n=\pm1$ [Fig.~\ref{fig:geometry}(a)]. We write
$s_n=\sin\alpha_n$ and $c_n=\cos\alpha_n$; the axial length $L_n$ satisfies
$L_n=s_nL_{n-1}$.

Let $\widehat{\Smat}^{(n)}$ denote the integrated $6\times6$ compliance that maps end forces and moments to end translations and rotations. The microscopic phase $\phi\in[0,2\pi)$ labels position within one helical turn. The $3\times3$ rotation $\mathsf Q_n(\phi)$ expresses the local filament directors in the outer-axis basis, while $\bm\rho_n=R_n\bm e_r(\phi)$ is the moment arm from that axis to the local centerline and $\bm e_r$ is the radial unit vector. The vectors $\bm F_n,\bm M_n$ are the force and moment resultants of the outer effective rod; $\bm f_{n-1},\bm m_{n-1}$ are the corresponding resultants resolved on the local level-$(n-1)$ precursor. Force and moment balance give
\begin{equation}
 \binom{\bm f_{n-1}}{\bm m_{n-1}}=
 \Tmat_n(\phi)\binom{\bm F_n}{\bm M_n},\quad
 \Tmat_n=
 \begin{pmatrix}
 \mathsf Q_n^T&\bm0\\[-1mm]
 -\mathsf Q_n^T[\bm\rho_n]_{\times}&\mathsf Q_n^T
 \end{pmatrix}.
 \label{eq:wrench}
\end{equation}
Equating complementary energy before and after coarse graining yields
\begin{equation}
 \widehat{\Smat}^{(n)}=
 \phaseavg{\Tmat_n^{T}(\phi)\widehat{\Smat}^{(n-1)}\Tmat_n(\phi)}.
 \label{eq:RG}
\end{equation}
Here $\Tmat_n$ is the $6\times6$ force--moment (wrench) transfer matrix, the
superscript $T$ denotes transpose, $\bm0$ is the $3\times3$ zero block,
$[\bm\rho]_{\times}\bm v=\bm\rho\times\bm v$ for any vector $\bm v$, and
$\phaseavg{A}=(2\pi)^{-1}\int_0^{2\pi}A(\phi)\,\dd\phi$ denotes the phase
average of any phase-dependent matrix $A(\phi)$. Coarse graining means
replacing the phase-resolved helix by a uniform rod with the same complementary energy. The factor $1/s_n$ in a compliance density is absorbed by the length relation in the integrated form. Equation~\eqref{eq:RG} is linear and preserves reciprocity and positive definiteness. Material properties specify only $\widehat{\Smat}^{(0)}$; geometry determines the flow. The map is exact within this periodically homogenized linear-rod model and becomes asymptotically exact for nested helices when every microscopic wavelength is small compared with the next deformation scale \cite{Gomez2025,Antman2005,Audoly2010,SM}. Finite-turn end layers, contact, and forming prestress are excluded \cite{Elata2004,Usabiaga2008,Liu2019}.

Phase averaging first restricts a general symmetric compliance to a seven-parameter transversely isotropic chiral form. For a circular isotropic precursor, two invariant relations reduce this form to the closed five-dimensional sector $\bm x_n=(a_n,b_n,d_n,e_n,q_n)^T$: transverse shear, axial extension, bending, torsion, and extension--twist coupling. ``Closed'' means that no additional compliance component is generated at later levels. The $5\times5$ transfer matrix in $\bm x_n=\Mmat_n\bm x_{n-1}$ is given in the Supplemental Material \cite{SM}. Its rotational sector diagonalizes through
\begin{equation}
 \mu_n=\frac{2d_n+e_n}{3},\qquad \Delta_n=d_n-e_n,
\end{equation}
The corresponding recursions are
\nopagebreak[4]
\begin{align}
 \mu_n&=\mu_{n-1},&
 \Delta_n&=\lambda_n\Delta_{n-1},\nonumber\\
 q_n&=\lambda_nq_{n-1}+\chi_nR_nc_ns_n\Delta_{n-1},&
 \lambda_n&=\frac{3s_n^2-1}{2}.
 \label{eq:rotsector}
\end{align}
Thus the mean rotational compliance is invariant, whereas the bending--torsion compliance contrast $\Delta_n=d_n-e_n$ is multiplied by a geometry-only factor. This contrast compares two deformation modes and does not imply that the base material is anisotropic. For identical levels, $\Mmat_n=\Mmat$ and $\lambda_n=\lambda$, and
\begin{equation}
 \det(z\I-\Mmat)=(z-1)^2(z-\lambda)^3.
 \label{eq:spectrum}
\end{equation}
Here $z$ is the spectral parameter and $\I$ is the $5\times5$ identity.
Generically, the eigenvalue $1$ has a size-two Jordan block and $\lambda$ has a size-three block. The spectrum is independent of Young's modulus, shear modulus, Poisson's ratio, filament radius, and length.

The marginal Jordan mode controls axial stiffness. With $m_n=(2a_n+b_n)/3$, the transfer law gives the exact identity
\begin{equation}
 m_n-m_{n-1}=\frac{R_n^2}{3}(d_{n-1}+e_{n-1}).
 \label{eq:mrec}
\end{equation}
This identity itself does not assume a radius-growth law.

We denote by $K_N=b_N^{-1}$ the torque-free axial stiffness of an $N$-level
hierarchy.

A physical nested centerline must additionally possess separated scales. Let $\Lambda_n=2\pi R_n/c_n$ be the wavelength of one level-$n$ turn and $\varepsilon_{n+1}=\Lambda_n/\Lambda_{n+1}$. Uniform separation, $\varepsilon_{n+1}\leq\varepsilon_*\ll1$ with an $N$-independent upper bound $\varepsilon_*$, together with angles in a compact subset of $(0,\pi/2)$ implies
$R_n/R_N=(c_n/c_N)\prod_{j=n+1}^{N}\varepsilon_j$ and hence $\sum_{n\leq N}R_n^2\asymp R_N^2$. The stable compliance recurrence then gives \cite{SM}
\begin{equation}
 K_N\asymp R_N^{-2},
 \label{eq:physicalclass}
\end{equation}
where $\asymp$ denotes equality of asymptotic order up to positive constants. This is the physical $N\to\infty$ class under uniform scale separation.

\begin{figure*}[t]
\includegraphics[width=0.95\textwidth]{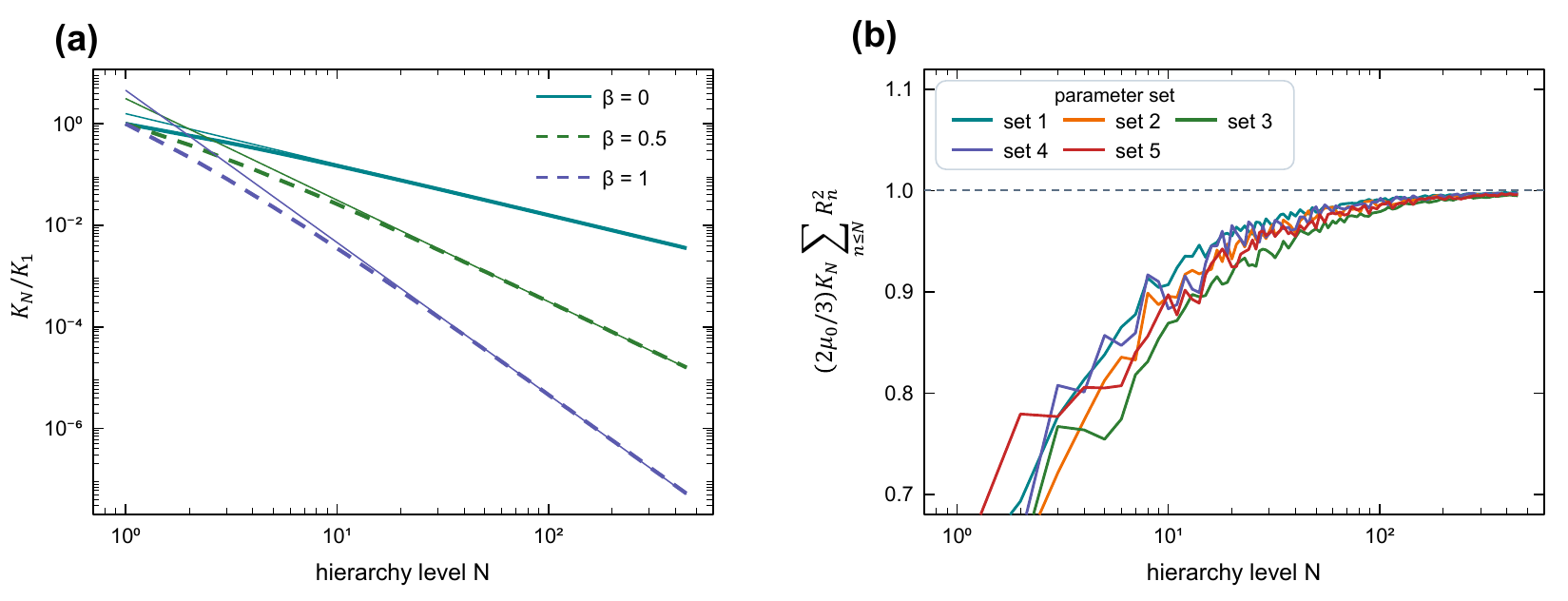}
\caption{\textbf{Renormalization-space stiffness classes.} (a) Log--log plots of normalized axial stiffness along regularly varying radius schedules in the homogenized recursion. Independent random helix angles retain the predicted slopes $-(2\beta+1)$ for $R_n\propto n^\beta$; thin lines show those powers. These trajectories expose the operator exponents but are not uniformly scale-separated infinite centerlines. (b) A parameter-free operator-level collapse: five choices of elastic constants, filament radius, and pitch disorder approach $(2\mu_0/3)K_N\sum_{n\leq N}R_n^2=1$. Uniformly scale-separated physical hierarchies instead obey Eq.~\eqref{eq:physicalclass}.}
\label{fig:scaling}
\end{figure*}

\begin{figure*}[t]
\includegraphics[width=0.95\textwidth]{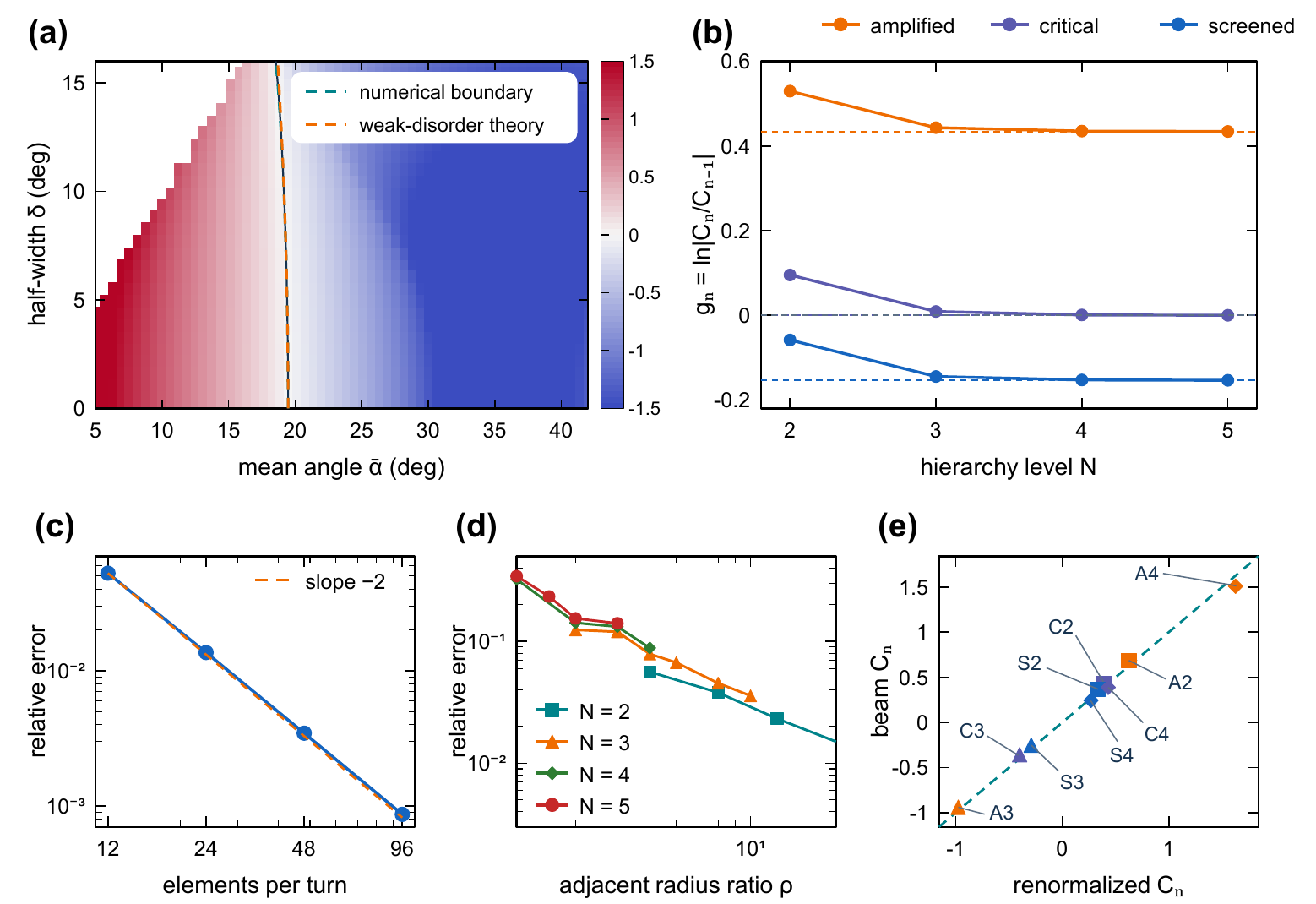}
\caption{\textbf{Finite-level chirality and independent validation.} (a) Lyapunov exponent $\gamma_\chi$ for angles uniformly distributed on $[\bar\alpha-\delta,\bar\alpha+\delta]$, where $\bar\alpha$ is the mean angle and $\delta$ the half-width; the solid contour is $\gamma_\chi=0$ and the dotted curve is the weak-disorder expansion. (b) One-step rate $g_N$ for deterministic amplified ($15^\circ$), asymptotically critical ($\alpha_c$), and screened ($21^\circ$) hierarchies with $\rho=10$; dashed lines are $\ln|y_\chi|$. (c) One-level beam convergence. (d) Full extension--twist matrix error through $N=5$ versus the common adjacent-radius ratio $\rho=R_{n+1}/R_n$. (e) Direct finite-centerline values of $\mathcal C_N$ versus renormalized predictions. Labels A, C, and S denote the three angle choices; $N=2,3$ use $\rho=10$, while the common noncontacting $N=4$ test uses $\rho=6$.}
\label{fig:chirality}
\end{figure*}

By contrast, regularly varying radius schedules are trajectories of the homogenized operator. Their radii obey $R_n=R_0n^\beta\ell(n)$, where $R_0>0$ is a radius scale, $\beta$ is the radius-growth exponent, and $\ell$ is slowly varying, meaning $\ell(tn)/\ell(n)\to1$ for every fixed $t>0$. For fixed or slowly varying angles $\varepsilon_{n+1}\to1$, and under bounded stationary disorder it is not uniformly small. They therefore do not represent uniformly separated infinite centerlines, although they can characterize formal renormalization classes or finite windows whose separation is checked independently. If $\beta>-1/2$ and $|\lambda_n|\leq\lambda_*<1$ for an $N$-independent bound $\lambda_*$, then
\begin{equation}
 K_N^{-1}\sim\frac{2\mu_0}{3}\sum_{n=1}^{N}R_n^2,
 \label{eq:masterstiff}
\end{equation}
and
\begin{equation}
 K_N\sim\frac{3(2\beta+1)}{2\mu_0R_0^2}
 \frac{N^{-(2\beta+1)}}{\ell(N)^2}.
 \label{eq:universal}
\end{equation}
Here $\sim$ denotes asymptotic equivalence, meaning that the ratio tends to
unity, whereas $\order(\cdot)$ denotes a term bounded in magnitude by a
constant multiple of its argument.
The condition $\beta>-1/2$ selects algebraic accumulation; the boundary case
is logarithmic. The material-independent exponent $2\beta+1$ reflects the prescribed moment-arm schedule. Figure~\ref{fig:scaling} tests these operator-level powers and collapse; Fig.~\ref{fig:chirality}(d) separately tests finite physical centerlines as adjacent scales are separated.

The integrated extension--twist coefficient $q_N$ has an immediate experimental interpretation. Let $\delta_N$ be the total end-to-end axial extension and $\Theta_N$ the relative end rotation about the final hierarchy axis. For axial force $F$ and torque $M$,
\begin{equation}
 \binom{\delta_N}{\Theta_N}=
 \begin{pmatrix}b_N&-q_N\\-q_N&e_N\end{pmatrix}
 \binom{F}{M}.
 \label{eq:endLaw}
\end{equation}
Thus a torque-free test measures $\Theta_N/(FL_N)=-q_N/L_N$. The
dimensionless normalized coupling
$\eta_N=q_N/\sqrt{b_Ne_N}$ obeys $|\eta_N|<1$ by positive definiteness, so
growth after rescaling does not imply mechanical instability.

Define $\widetilde\Delta_n=\Delta_n/L_n$ and $\widetilde q_n=q_n/L_n$. Their exact recursion is
\begin{align}
 \widetilde\Delta_n&=y_{\chi,n}\widetilde\Delta_{n-1},\nonumber\\
 \widetilde q_n&=y_{\chi,n}\widetilde q_{n-1}
 +\chi_nR_nc_n\widetilde\Delta_{n-1},\nonumber\\
 y_{\chi,n}&=\frac{\lambda_n}{s_n}
 =\frac{3\sin^2\alpha_n-1}{2\sin\alpha_n}.
 \label{eq:chiralfield}
\end{align}
For a stationary ergodic angle sequence, the homogeneous mode has Lyapunov exponent \cite{Furstenberg1960,Oseledets1968}
\begin{equation}
 \gamma_\chi=\lim_{N\to\infty}\frac1N
 \ln\left|\frac{\widetilde\Delta_N}{\widetilde\Delta_0}\right|
 =\mathbb E\!\left[\ln|y_\chi(\alpha)|\right].
 \label{eq:lyap}
\end{equation}
Here $\mathbb E$ denotes expectation over the stationary distribution of the
helix angle $\alpha$, and $y_\chi(\alpha)$ is the multiplier in
Eq.~\eqref{eq:chiralfield} evaluated at that angle. Uniform scale separation
makes the physical radii grow
at least geometrically. Let
$\kappa_R=\lim_{N\to\infty}N^{-1}\ln(R_N/R_1)>0$ when this limit exists. Under the homochiral, noncanceling source conditions stated below, the unnormalized measurable density has exponent $\gamma_\chi+\kappa_R$. Removing this deterministic moment-arm growth defines the outer-radius-normalized observable
\begin{equation}
 \mathcal C_N=\frac{\widetilde q_N}{R_N}
 =-\frac{\Theta_N}{FL_NR_N},\qquad
 \Gamma_N^{(\mathcal C)}=\frac1{N-1}\ln\left|\frac{\mathcal C_N}{\mathcal C_1}\right|.
 \label{eq:normalizedObservable}
\end{equation}
Here $\Gamma_N^{(\mathcal C)}$ is the cumulative per-level growth rate of
$\mathcal C_N$. For a homochiral hierarchy, meaning $\chi_n=\chi$ for every
$n$, whose angle support remains on one fixed-sign branch of $y_\chi$ and is
separated from $y_\chi=0$, and with nonzero precursor contrast
$\widetilde\Delta_0$, the exact source solution and
$R_k/R_N\le C_{\rm src}\varepsilon_*^{N-k}$ for an $N$-independent constant
$C_{\rm src}>0$ give
$\Gamma_N^{(\mathcal C)}\to\gamma_\chi$ almost surely \cite{SM}. Hence the stiffness law $K_N\asymp R_N^{-2}$ and the measurable chirality threshold $\gamma_\chi=0$ coexist in the same uniformly scale-separated physical hierarchy. For a periodic hierarchy, $|y_\chi|=1$ gives $\alpha_c=\sin^{-1}(1/3)=19.47^\circ$; $\alpha_*=\sin^{-1}(1/\sqrt3)=35.26^\circ$ instead sets $\lambda=0$ and erases the bending--torsion compliance contrast in one step [Fig.~\ref{fig:geometry}(c)].

The cumulative rate can converge slowly near $\gamma_\chi=0$, but a finite hierarchy admits a faster precursor. For constant $\alpha$, geometric radii $R_n/R_{n-1}=\rho>1$, where $\rho$ is the common adjacent-radius ratio, and an uncoupled base filament with $\widetilde\Delta_0\ne0$, define the measurable one-step rate $g_N=\ln|\mathcal C_N/\mathcal C_{N-1}|$. The exact source sum gives \cite{SM}
\begin{equation}
 g_N=\ln|y_\chi|+
 \ln\frac{1-\rho^{-N}}{1-\rho^{-(N-1)}}.
 \label{eq:finitegain}
\end{equation}
The correction is geometric rather than $\order(N^{-1})$: at $\rho=10$ it is $9.05\times10^{-3}$ already at $N=3$. Figure~\ref{fig:chirality}(b) shows that three levels distinguish representative amplified and screened sequences. The companion finite stiffness exponent $p_N=-\ln(K_N/K_{N-1})/\ln\rho$ differs from $2$ by at most $3.71\times10^{-2}$ at $N=3$ for the same three sequences.

Pitch disorder shifts rather than removes the threshold. For angles uniformly
distributed on $[\bar\alpha-\delta,\bar\alpha+\delta]$, with mean
$\bar\alpha$ and half-width $\delta$, the threshold curve is
\begin{equation}
 \frac{1}{2\delta}\int_{\bar\alpha-\delta}^{\bar\alpha+\delta}
 \ln\left|\frac{3\sin^2\alpha-1}{2\sin\alpha}\right|\dd\alpha=0.
 \label{eq:disorderboundary}
\end{equation}
Weak disorder gives
$\bar\alpha_c(\delta)=\alpha_c-(\sqrt2/8)\delta^2+\order(\delta^4)$, where
$\bar\alpha_c$ is the critical mean angle and all angles are in radians.
Figure~\ref{fig:chirality}(a) shows the full threshold curve. Near a regular
crossing, the hierarchy crossover depth $N_\times=1/|\gamma_\chi|$ diverges
with the material-independent exponent one; $N_\times$ is the number of
levels required for an order-one amplification or attenuation. For
disordered finite samples, averaging the measured step rates over specimens
or level blocks estimates the expectation in Eq.~\eqref{eq:lyap}.

A finite-depth experiment can determine $g_N$ from torque-free tensile tests
on matched $(N-1)$- and $N$-level specimens by measuring
$\Theta_N/(FL_NR_N)$ [Fig.~\ref{fig:geometry}(b)]. At $\rho=10$, recursion
gives $N=3$ gains $1.558$ at $15^\circ$ and $0.865$ at $21^\circ$; the
direct centerline values are $1.37$ and $0.679$, preserving the amplified
and screened classifications. Thus $5$--$10\%$ relative precision separates
these representative cases. At the asymptotic critical angle, however, the
direct finite-scale gain is $0.825$, so locating the finite-depth crossing
requires an angle sweep or calibration rather than a nominal $1\%$ test at
$\alpha_c$. The test must remain linear, slender, scale separated, and
noncontacting; no modulus calibration is required for the ratio.

We tested the theory without experimental fitting using the actual nested
centerline, represented by straight three-dimensional Euler--Bernoulli beam
segments. Direct complementary-energy integration uses neither phase
averaging nor the reduced transfer matrix and agrees with a global sparse
stiffness implementation through a representative three-level geometry to
$3\times10^{-8}$ in the full matrix. One-level mesh error converges
quadratically [Fig.~\ref{fig:chirality}(c)]. The full $2\times2$
extension--twist matrix approaches the renormalized result through five
levels as scale separation increases [Fig.~\ref{fig:chirality}(d)];
representative errors are $0.47\%$, $3.6\%$, $8.9\%$, and $14.1\%$ at
$N=2,3,4,5$. More stringently, one common $N=4$ geometry with $\rho=6$ and
24 elements per innermost turn preserves the sign of $\mathcal C_N$ for all
three angle choices, with magnitude errors $7.2$--$10.0\%$ and full-matrix
errors $0.75$--$2.32\%$ [Fig.~\ref{fig:chirality}(e)]. Exact
segment--segment screening of the discretized centerline gives a minimum
clearance of $3.51$ filament diameters. Five-level chiral calculations remain
parameter-dependent diagnostics and are not used for quantitative
validation \cite{SM}.

Han \textit{et al.} resolve stiffness, contact, and tension--torsion coupling
or decoupling in prescribed finite hierarchies \cite{Han2026}. Here
finite-depth decoupling and Lyapunov screening are not synonymous. In the
present notation, finite decoupling is a condition such as $q_N=0$, or a
strongly suppressed off-diagonal compliance, for a selected finite
construction. Screening instead means $\gamma_\chi<0$ for the
outer-radius-normalized response $\mathcal C_N$: it is an exponential decay
rate with hierarchy depth and does not require $q_N$ to vanish at any finite
level. Our distinct results are the spectrum and marginal Jordan sector of
the iterated arbitrary-precursor map, its radius law and Lyapunov threshold,
and a finite-step statistic visible at small depth.

Recursive coiling therefore changes the question from the response of a prescribed construction to the dynamics of an iterated constitutive map. A marginal compliance mode produces an accumulated-radius law: uniform physical scale separation selects the outer-radius inverse-square class, while regularly varying schedules expose intermediate operator exponents. After deterministic outer-radius growth is removed, a scalar multiplicative mode produces a disorder-robust chirality threshold. The finite-step rate makes its approach measurable by three levels for representative sequences rather than only in an infinite hierarchy. These results are independent of the base material because elasticity supplies only the initial point of a geometry-only flow. Dynamics, finite deformation, prestress, and contact lie beyond this solvable noncontacting fixed point.

\section*{End Matter}
\textit{Scale separation and finite-level remainder.--}
Equation~\eqref{eq:RG} follows by inserting Eq.~\eqref{eq:wrench} into the complementary energy of the level-$(n-1)$ rod and differentiating with respect to the outer end resultant. With $\varepsilon_{n+1}=\Lambda_n/\Lambda_{n+1}$,
$R_n/R_N=(c_n/c_N)\prod_{j=n+1}^{N}\varepsilon_j$. Thus $\varepsilon_j\leq\varepsilon_*<1$ and bounded angles imply
$R_N^2\leq\sum_{n\leq N}R_n^2\leq C_{\rm sep}R_N^2/(1-\varepsilon_*^2)$
for an $N$-independent geometric constant $C_{\rm sep}>0$; the stable
compliance recurrence then yields Eq.~\eqref{eq:physicalclass}. In the
isotropic circular sector,
$\Delta_N=\Delta_0\prod_{j\leq N}\lambda_j$ exactly. Under the
regular-variation assumptions,
\begin{equation}
 K_N^{-1}=\frac{2\mu_0}{3}\sum_{n\leq N}R_n^2
 +\order(R_N^2)+\order\!\left(\sum_{n\leq N}R_n^2
 \prod_{j<n}|\lambda_j|\right),
 \label{eq:remainder}
\end{equation}
and both remainders are subleading, proving Eq.~\eqref{eq:masterstiff}. The complete arguments and the distinction between the physical and operator limits are in the Supplemental Material \cite{SM}.

\textit{Numerical protocol.--}
The direct beam calculation uses the Kirchhoff specialization of the level-$0$ rod, exact rigid-arm transfer of terminal force and torque to the outer axis, and complementary-energy integration along straight centerline segments. A global six-degree-of-freedom-per-node sparse stiffness calculation provides an implementation cross-check. Centerlines through five levels are generated by composing local director frames of exact circular helices. Mesh, outer-turn, scale-separation, and segment-clearance checks accompany the data. Initial stress is absent because both theories linearize about the same prescribed stress-free centerline. No parameter is fitted to the recursive result.

\end{document}